# A solution for the differences in the continuity of continuum among mathematicians

Haidong. Zhu[*]

*Department of Engineering Mechanics, Hohai University, Nanjing, Jiangsu 210098, China*

**Abstract** There are the longstanding differences in the continuity of continuum among mathematicians. Starting from studies on a mathematical model of contact, we construct a set that is in contact everywhere by using the original idea of Dedekind's cut and weakening Order axioms to violate Order axiom 1. It is proved that the existence of the set constructed can eliminate the differences in the continuity.

## 1 Introduction

The continuity of continuum, as we know, characterizes the unity of continuum (cf. Hegel 1928; Dedekind 1963). However, there are longstanding differences in the continuity of continuum among mathematicians. Summarily two different types of continuity are held, called the *Cantor-type continuity* and the *Poincaré-type continuity* for distinction in this paper.

The mathematicians represented by Cantor and Dedekind, suggest that a point set has the continuity if it is complete (e.g., Yu. et al. 2010; Fernández and Thiagarajan 1984; Lücke. et al. 1999; Bergé 2008). We call the continuity that is defined by the completeness, the *Cantor-type continuity*. The development of continuum concept has been experiencing a process that gradually evolves and increasingly improves the structure of number systems. After the appearances of natural numbers $\mathbb{N}$, the integral numbers $\mathbb{Z}$, the rational numbers $\mathbb{Q}$ in succession, finally, the real numbers $\mathbb{R}$ is constructed from the rational numbers $\mathbb{Q}$. An elegant method to arrive at this goal is Dedekind cuts in which one visualizes real numbers as places at which a line may be cut with scissors (see Dedekind 1963; Bauer and Taylor 2009; Reck 2003; Schuster 2003). According to the theory of Dedekind cuts, there are some essential differences between the rational numbers $\mathbb{Q}$ and the real numbers $\mathbb{R}$. Typically, $\mathbb{R}$ has the "Cantor-type continuity" by the theorem as follows

> The set $\mathbb{R}$ constructed by means of Dedekind cuts, is (Cantor-type) continuous (i.e., complete) in the sense that it satisfies the Least Upper Bound Property: if S is a non-empty subset of $\mathbb{R}$ and is bounded above then in $\mathbb{R}$ there exists a least upper bound for S.

---

[*]*E-mail addresses:* cute39@163.com (H.D. Zhu)

By continuous we mean intuitively that there are no "gaps" between real numbers. In the rational numbers, by contrast, there is intuitively a gap where each of the numbers such as $\sqrt{2}$, $\pi$ and $e$ ought to be, but are not.

However, from intuitionists, it is considered unlikely that all distinct real points are linked to make a true continuous line. The intuitionists like Aristotle, Brouwer, Weyl (1918) and their followers all shared the same view that the continuum can by no means be regarded as merely a collection (set) of distinct elements (points). Poincaré, as typical representative of intuitionists, held to the viewpoint: since distinct numbers do not touch (connect) each other [or in other words, they are mutually disjoint (external to each other)], then real numbers (position points of the number axis) can not yield a continuum (cf. Boyer 1959). It is the famous Poincaré's remark to be mentioned in Russell's classic book, The Principles of Mathematics (Russell 1937, p.347), and it reads as follows

> *The continuum thus conceived is nothing but a collection of individuals arranged in a certain order, infinite in number, it is true, but external to each other. This is not the ordinary concept, in which there is supposed to be, between the elements of the continuum, a sort of intimate bond which makes a whole of them, in which the point is not prior to the line, but the line to the point. Of the famous formula $2^{\aleph_0} = c$, the continuum is unity in multiplicity, the multiplicity alone subsists, the unity has disappeared.*

Clearly Poincaré's remark shows that $\mathbb{R}$ is an ordered set of infinitely many elements (points) having all its elements external to each other, namely, all the real points (numbers) in $\mathbb{R}$ are disconnected. But, based on Poincaré's idea, the existence of "intimate bond" can make the real points are not external to each other and continuous. We call the continuity that is defined by the "intimate bond", the *Poincaré-type continuity*. Regrettably, the Poincaré-type continuity has not been developed well in theory because of the lack of excellent mathematical model.

This paper points out the underlying cause of the above differences is that the real numbers $\mathbb{R}$ satisfies Order axioms, more exactly, Order axiom 1 (i.e., Trichotomy Law) (see Palmgren 2005). According to Order axiom 1, all real points are necessarily disjoint and external to each other. But it is for that very reason that the mathematical model of the Poincaré-type continuity is hardly constructed under Order axioms.

It is easy to find that the Cantor-type continuity and the Poincaré-type continuity are merely a single-phase abstraction of the real continuity. Based on the characters of continuous variation in reality, the Cantor-type continuity characterizes no "gaps" in an object or a set; and, the Poincaré-type continuity characterizes that the distinct elements of an object (or a set) have a certain degree of correlation that bridges the distinct elements to make the object (set) is connected indeed. A conventional chain, for example, has the Poincaré-type continuity because the chain rings as the elements of the chain are not external to each other (namely, there are the correlation in the form of embedding or overlapping between adjacent chain rings). Intuitively, the

Cantor-type continuity reflects the compression-resisting property because of no "gaps" between elements, and the Poincaré-type continuity reflects the tension-resisting property because of the correlation between elements. It follows, therefore, that the Cantor-type continuity and the Poincaré-type continuity can not fully reflect the real continuity that characterizes the deformation-resisting property, which really make a whole of distinct elements consisting of capable of resisting compression and tension. Then, $\mathbb{R}$ is not Poincaré-type continuous without the spatial "correlation" between the real points, and a chain is not Cantor-type continuous with the embedding "gaps" between the adjacent chain rings.

To construct a set consisting of the real continuity as the synthesis of the Cantor-type continuity and the Poincaré-type continuity, we first introduce a novel mathematical model of contact by weakening Order axioms to violate Order axiom 1. By the contact model, we construct a set $W$ that is complete and in contact everywhere by using the original idea of Dedekind's cut (Dedekind 1963). According to Poincaré's idea, the contact model of this paper is just considered as the mathematical model of Poincaré's "intimate bond". It is proved that the set $W$ simultaneously has the Cantor-type continuity and the Poincaré-type continuity, in other words, the existence of $W$ can eliminate the longstanding differences in the continuity of continuum. Furthermore, using $W$ as a mathematical model to describe motion trajectory of an object, we can understand and describe accurately the motion continuity of an object, namely the moving state of an object from one position to another.

**2 Contact**

For a concrete object consisting of unity and continuity, its arbitrary two adjacent parts are in contact. Generally a nonempty bounded closed set is chosen as a mathematical model of a concrete object, because any concrete object has its own boundary by default. That is, the model of contact between adjacent parts of an object is mathematically that the distance of two nonempty bounded closed sets equals zero (see e.g., Constantinescu et al. 2005; Asarin et al. 2010). However, the common plausible mathematical model of contact contradicts the existing mathematical theories (essentially, Order axiom 1). To solve the contradiction, we first discuss the superposition of sets to introduce a new kind of sets, which does not necessarily meet Order axiom 1.

*2.1 Superposition*

By the classical theory of the mathematical analysis, the real numbers $\mathbb{R}$ has its own axiom system (Xiao 2008). Here, we list Order axiom 1 (Trichotomy Law) of $\mathbb{R}$ as follows:

**O1** *For two elements a, b in a set X, exactly one of a < b, a = b, a > b is true.*

O1 shows that arbitrarily equal-value elements must be identical, while arbitrarily unequal-value elements must be distinct. This clearly causes a reduction of elements' number of sets in superposition problems. For example, two sets $A = \{a, b, c\}$ and $B = \{b, c, d\}$ are superposed since their intersection $\{b, c\}$ is nonempty. Note that, $A$ and $B$ together have six elements but the union of $A$ and $B$ only contains four elements, i.e., $A \cup B = \{a, b, c, d\}$. The reason is that the elements $b$ and $c$ in $A$ is considered to have no distinction with the elements $b$ and $c$ in $B$ according to O1.

To prevent the reduction of elements in superposition problems, we use subscript to represent the belonging of elements. Then the equal-value elements can be distinguished by their subscripts. For example, $A = \{a, b, c\}$ can be denoted as $\{a_A, b_A, c_A\}$, and similarly $B = \{b, c, d\}$ can be denoted as $\{b_B, c_B, d_B\}$. Clearly, $\{b_A, c_A\} \neq \{b_B, c_B\}$ because the belonging of $\{b_A, c_A\}$ and $\{b_B, c_B\}$ are distinct. Then the union of $A$ and $B$ can contain six elements instead of four elements, i.e., $A \cup B = \{a_A, b_A, c_A, b_B, c_B, d_B\}$. Actually the ubiquitous superposition principle in scientific researches reflects the superposition that elements are not reduced (e.g., Johansson 1999; Hofheinz 2009). Thus, using subscript to represent the belonging of element is necessary and feasible in superposition problems.

For an ordered set $X$, we denote the subsets of $X$ by $X_i$ (where the subscripts $i$ are used to distinguish the different subsets of $X$), and the elements of $X_i$ by $x_i$ ($x \in X$). The subscripts $i$ of $x_i$ indicates obviously $x_i \in X_i$. We use the elements of certain (not necessarily all) different subsets $X_y$ ($y \in Y$) of $X$ to define a set $Z$ as follows

$$Z = \{x_y \mid x \in X, y \in Y\},$$

Here we call the set $X$ the *value source* of $Z$, denoted by $\underline{Z} = X$, and call the set $Y$ the *series source* of $Z$, denoted by $Z_\# = Y$. Writing $z = x_y$ ($z \in Z$), we call $x$ the *value* of $z$, denoted by $\underline{z} = x$, and call $y$ the *series* of $z$, denoted by $z_\# = y$. It must be particularly pointed out that, by the definition of $Z$, we have $Z = \bigcup_{y \in Y} X_y$. Then, for different subsets $X_y$, the subscripts $y$ are also different, and then, the set $Y$ that is composed of all $y$ can be chosen to be an ordered set based on Order axioms. This paper defines that the set $Y$ is always ordered. So the value source and the series source of $Z$ are ordered sets. Moreover, intersections of different subsets $X_y$ may be nonempty under Order axioms, because the superposition that satisfies O1 can reduce elements' number of the union of different sets $X_y$.

Now we discuss some properties of the set $Z$ in superposition that dissatisfies O1. We weaken Order axioms to violate O1 for the superposition of sets. This causes the intersection of any two different subsets $X_y, X_{y'} \subset Z$ of $X$ is empty (i.e., $X_y \cap X_{y'} = \varnothing$), in other words, there may exist $x_y \in X_y$, $x_{y'} \in X_{y'}$, then $x_y \neq x_{y'}$. We call two elements $u, v \in Z$ are *equal-value* (or *unequal-value*) if $\underline{u} = \underline{v}$ (or $\underline{u} \neq \underline{v}$), or *equal-series* (or *unequal-series*) if $u_\# = v_\#$ (or $u_\# \neq v_\#$). We define equality of $u, v \in Z$ according to $u = v$ if $\underline{u} = \underline{v}$ and $u_\# = v_\#$, and inequality of $u, v$ according to $u \neq v$ if $\underline{u} \neq \underline{v}$ or $u_\# \neq v_\#$. Note that the relations "<" and ">" are difficult to temporarily identify for all elements of

$Z$.

**Corollary 2.1** *In Z, the equal-value relation is an equivalence relation.*

**Proof** Let ~ be an equal-value relation in $Z$. For $u, v \in Z$, we have $u \sim v$ if $\underline{u} = \underline{v}$.
(1) (Reflexivity) for any $u \in Z$, $\underline{u} = \underline{u}$, that is, $u \sim u$.
(2) (Symmetry) when $\underline{u} = \underline{v}$, then, $\underline{v} = \underline{u}$. It implies $u \sim v, v \sim u$.
(3) (Transitivity) when $\underline{u} = \underline{v}$, $\underline{v} = \underline{w}$ ($w \in Z$), then, $\underline{u} = \underline{w}$. It implies $u \sim v, v \sim w, w \sim u$.

Therefore, ~ is the equivalence relation in $Z$. □

Similarly, we can prove that the equal-series relation is also equivalent.

**Definition 2.1** A set is *disordered* if it dissatisfies O1.

Since the value of an element represents the position of the element or the Cartesian coordinates of the element, the value $x \in X$ is just the position of element $z \in Z$. Note that, the series $z_\#$ only represents the belonging of $z$, but contains no position information of $z$. Thus in the problems involving spatial position relation of elements, a decision on whether the set $Z$ is disordered can only be taken by values of elements, not series of elements. Below, a set is considered to be disordered in the sense that there are some distinct elements that are equal-value, which is "forbidden" by O1.

**Theorem 2.2** *Let $Z = \{x_y \mid x \in X, y \in Y\}$. $Z$ is disordered if there exist two distinct elements $u, v \in Z$, then $\underline{u} = \underline{v}$.*

**Proof** By O1, for $u, v \in Z$, $\underline{u} = \underline{v}$ means $u = v$, which is contradictory to the known condition that $u, v$ are distinct, i.e., $u \neq v$. Then, $Z$ dissatisfies O1. By Definition 2.1, $Z$ is disordered. □

**Corollary 2.3** *Let $Z = \{x_y \mid x \in X, y \in Y\}$. $Z$ is ordered if there exists a one-to-one correspondence between $X$ and $Y$.*

**Proof** Let $f: x \to y$ ($x \in X, y \in Y$) be a one-to-one correspondence. Writing $y = f(x)$, $z = x_y$ ($z \in Z$) can be replaced by $z = x_{f(x)}$. Since $f$ is one-to-one, $z = x_{f(x)}$ implies $Z$ is in one-to-one correspondence with $X$; since $X$ is ordered, so that $Z$ is also ordered. □

By Corollary 2.3, the set $\mathbb{R}' = \{r_r \mid r \in \mathbb{R}\}$ is ordered; since the value source of $\mathbb{R}'$ is $\mathbb{R}$, $\mathbb{R}'$ is complete. Since $\mathbb{R}$ is, up to isomorphism, the only complete ordered field (Xiao 2008), $\mathbb{R}'$ is equivalent to $\mathbb{R}$. In general, we make no any distinction between $\mathbb{R}'$ and $\mathbb{R}$.

*2.2 Mathematical model*

Just as we know it, the sufficient and necessary condition of contact between any two objects from the perspective of space is that the distance of the two objects is zero. Generally two objects in contact are regarded as two nonempty bounded closed sets that are zero-distance. For example, we cut a finite length straight bar $l$ into two parts $l_1$ and $l_2$, which are also finite length straight bars. Taking a closed line segment as mathematical model of a finite length straight bar, $l$, $l_1$ and $l_2$ are all nonempty bounded closed sets. Then we have naturally, $l_1 \cup l_2 = l$, $l_1 \cap l_2 = \varnothing$, and $\rho(l_1, l_2) = 0$ [where $\rho(l_1, l_2)$ means the distance between $l_1$, $l_2$]. That is to say, there exist two zero-distance disjoint line segments (nonempty bounded closed sets). This is in contradiction with the existing set theory in which the following theorem is explicitly given: the distance between any two nonempty bounded closed sets is always positive (see Borodich and Feng 2010; Tkachuk 1992).

In fact, the essential reason that causes the above contradiction is that the mathematical models of concrete objects are constructed in the categories of Order axioms, more exactly, Order axiom 1 (i.e., Trichotomy Law) (cf. Asarin et al. 2010; Haslinger et al. 2009). Since for the two parts $l_1$, $l_2$ of the straight bar $l$, $l_1 \cap l_2 = \varnothing$, then we have $s \neq t$ for any two points (elements) $s \in l_1$, $t \in l_2$. In the categories of Order axioms, $s \neq t$ inevitably causes $\rho(s, t) > 0$, and then $\rho(l_1, l_2) > 0$ instead of $\rho(l_1, l_2) = 0$ according to the definition of distance between sets. That is, the contact between two disjoint closed line segments, as mathematical models of two straight bars, is impossible to appear theoretically, which is a contradiction to the physical reality.

So we use a disordered nonempty bounded closed set to describe an object in contact problems. Correspondingly, we use value of element to represent position of element of the object. Thus the theory concerning closed set can be used directly to study the contact problems.

We say that a point (element) $u$ is a *boundary point* of a set $Z$ if every neighborhood of $u$ contains both points belonging to $Z$ and points not belonging to $Z$. The set of boundary points of $Z$ forms the *boundary* of $Z$, denoted by the symbol, $\partial Z$. Generally a set $Z$ is *closed* if it contains its boundary. For $u, v \in Z$, we denote their values $\underline{u} = (\underline{u}_1, \underline{u}_2, ..., \underline{u}_n)$ and $\underline{v} = (\underline{v}_1, \underline{v}_2, ..., \underline{v}_n)$, where the numbers $\underline{u}_n$, $\underline{v}_n$ are the Cartesian coordinates of $u, v$. The distance of $u$ and $v$, denoted by $\rho(u, v)$, is given by

$$\rho(u, v) = \left[\sum_{i=1}^{n}(\underline{u}_n - \underline{v}_n)^2\right]^{1/2},$$

and the distance of two nonempty sets $A$ and $B$, denoted by $\rho(A, B)$, is given by

$$\rho(A, B) = \inf\{\rho(u, v) \mid u \in A, v \in B\}.$$

Based on the above theory of closed set, we can obtain easily the theorem as follows.

**Theorem 2.4** *Let A, B be two nonempty bounded closed sets. Then $\rho(A, B) > 0$ if $\underline{A} \cap \underline{B} = \varnothing$.*

Since objects are nonempty bounded closed in the contact problems, the conditions of Theorem 2.4 can be met naturally in the contact problems. Actually, the condition "$A$, $B$ are two nonempty bounded closed sets" can be relaxed to "$A$ is a nonempty bounded closed set and $B$ is a nonempty closed set". That is, at least one of $A$ and $B$ is bounded. But the condition can not be relaxed to "$A$, $B$ are two nonempty closed sets". Note that Theorem 2.4 can be used in the category satisfying or dissatisfying O1. In the category satisfying O1, the condition "$\underline{A} \cap \underline{B} = \varnothing$" can be reduced to the condition "$A \cap B = \varnothing$".

In the contact problems, for two disjoint objects $A$, $B$ as nonempty bounded closed sets, the sufficient and necessary condition of contact is $\rho(A, B) = 0$. That is, the contact phenomena in reality cause inevitably the existence of zero-distance disjoint nonempty bounded closed sets in theory, which is a contradiction to O1 according to Theorem 2.4. Therefore, the mathematical model of contact is disordered essentially.

**Definition 2.2** Two disjoint objects $A$, $B$ are *in contact* if $\exists\, a \in A, b \in B$, then $\underline{a} = \underline{b}$.

For example, with value sources $\underline{A} = [0, 1]$ and $\underline{B} = [1, 2]$, two sets $A$, $B$ are in contact since $\underline{A} \cap \underline{B} = \{1\} \neq \varnothing$, i.e., $\rho(A, B) = 0$. We denote the contact points of $A$, $B$ by $1_A$, $1_B$, respectively. Then, $1_A \neq 1_B$, and $\underline{1}_A = \underline{1}_B$. This shows the distance of two nonempty bounded closed sets may be zero indeed.

**Theorem 2.5** *The following are equivalent, for two objects A and B:*
  (i) *A and B are in contact.*
  (ii) $\underline{A} \cap \underline{B} \neq \varnothing$.
  (iii) $\rho(A, B) = 0$.

**Proof** (i) $\to$ (ii). Two objects $A$, $B$ are in contact in the sense that, $\exists\, a \in A, b \in B$, then, $\underline{a} = \underline{b}$, and then $\underline{A} \cap \underline{B} \neq \varnothing$.

(ii) $\to$ (iii). Since $\underline{A} \cap \underline{B} \neq \varnothing$, then $\exists\, a \in A, b \in B, \underline{a} = \underline{b}$, and then the distance $\rho(a, b) = 0$. Since the distance $\rho(A, B) = \inf\{\rho(u, v) \mid u \in A, v \in B\}$, then $\rho(A, B) \leqslant \rho(a, b)$, that is, $\rho(A, B) = 0$.

(iii) $\to$ (i). As two objects, $A$, $B$ are nonempty bounded closed. Without loss of generality, suppose that $\underline{u} \leqslant \underline{v}$ for any two elements $u \in A, v \in B$. Let $\underline{a} = \max\{\underline{u}\}$ and $\underline{b} = \max\{\underline{v}\}$ ($a \in A, b \in B$), then $\underline{a} \leqslant \underline{b}$, and then $\rho(A, B) = \rho(a, b)$. Since $\rho(A, B) = 0$, then $\rho(a, b) = 0$, i.e., $\underline{a} = \underline{b}$. By Definition 3.1, $A$, $B$ are in contact. $\square$

### 3 Eliminating differences

*3.1 Dedekind-type cut*

By making use of the original idea of Dedekind's cut, we will construct a disordered set, to eliminate the differences in the continuity. Recall that every Dedekind's cut of rational numbers, say $(A \mid B)$ with non-empty $A$ and $B$ so that $A \cup B = \mathbb{Q}$ (the set of rational numbers) and for any $a \in A, b \in B, a < b$, just defines a real number of $\mathbb{R}$.

**Definition 3.1** Let $W$ ($\underline{W} = \mathbb{R}$) be a point set. A *Dedekind-type cut* $(A \mid B)$ of $W$ is a partition of $W$ into two subsets $A$ and $B$, such that
   a. $A \cup B = W, A \neq \varnothing, B \neq \varnothing, A \cap B = \varnothing$;
   b. if whenever $a \in A, b \in B$, then $\underline{a} \leqslant \underline{b}$.

**Types of Dedekind-type Cuts** There are three possible types of Dedekind cuts $(A \mid B)$ of $W$, for we may have the situations where
   (1) *A has a largest value point and B has no smallest value point.*
   (2) *A has no largest value point and B has a smallest value point.*
   (3) *A has a largest value point and B has a smallest value point.*

A seeming fourth possibility, where $A$ has no largest value point and $B$ has no smallest value point, can not occur, for between any two distinct real numbers $\underline{a} \in \underline{A}$, $\underline{b} \in \underline{B}$ there is always a real number $\underline{c}$ ($c \in W$) could not lie in either $\underline{A}$ and $\underline{B}$, that is, $c \notin A$ and $c \notin B$, which contradicts the formation of the sets $A$ and $B$.

A Dedekind-type cut that is of type (1) or (2) is called an *ordered cut*. By the original idea of Dedekind's cut, the point set $W$ defined by the ordered cut is ordered and complete, which up to isomorphism is equivalent to $\mathbb{R}$. So the following we only use $W$ to denote the point set defined by the Dedekind-type cut that is of type (3).

Before discussing the Dedekind-type cut that is of type (3), we first give the definition of the Poincaré-type continuity from the viewpoint of continuity of intuitionists. According to Poincaré's remark, our contact model, as a correlation between adjacent distinct sets, is just considered as an idealization of Poincaré's "intimate bond", because the adjacent distinct sets are indeed connected to each other with the existence of equal-value elements. Thus, we define the Poincaré-type continuity differing from the Cantor-type continuity as follows.

**Definition 3.2** A set $Z$ is *Poincaré-type continuous*, if any two subsets $A$ and $B$ are in contact, where $A \cup B = Z, A \neq \varnothing, B \neq \varnothing$, and $A \cap B = \varnothing$.

By using the contact model as an idealization of Poincaré's "intimate bond", a Poincaré-type continuous set is in contact everywhere. Then by Definition 3.3, for the Dedekind-type cut that is of type (3), we can obtain the following conclusions.

**Theorem 3.1** Let $(A \mid B)$ be a Dedekind-type cut of $W$ ($\underline{W} = \mathbb{R}$). *W is Poincaré-type continuous (i.e., in contact everywhere), if A has a largest value point and B has a smallest value point.*

**Proof** Let $a$ be the largest value point of $A$, and let $b$ be the smallest value point of $B$. By Definition 3.1, we have $\underline{a} \leqslant \underline{b}$, then $\underline{u} \leqslant \underline{a}, \underline{b} \leqslant \underline{v}$ for any $u \in A, v \in B$.

Suppose that $\underline{a} \neq \underline{b}$, i.e., $\underline{a} < \underline{b}$. Then there is a point $c$ such that $\underline{c} = (\underline{a} + \underline{b})/2$. Since $\underline{a} < \underline{c} < \underline{b}$, it follows that $c \notin A$ and $c \notin B$, which contradicts that $\underline{A} \cup \underline{B}$ ($= \underline{W} = \mathbb{R}$) is complete. Thus, we have $\underline{a} = \underline{b}$. By Definition 3.1, $A$ and $B$ are in contact; and then $W$ is in contact everywhere (Poincaré-type continuous) according to Definition 3.2. □

According to the proof of Theorem 3.1 and Definition 2.1, the point set $W$ defined by the Dedekind-type cut that is of type (3) is disordered. Here we call the Dedekind-type cut that is of type (3) the *disordered cut*.

Theorem 3.1 shows that $W$ defined by the disordered cut has essential differences from $\mathbb{R}$. By the disordered cut, in $W$ the largest value point $a$ of $A$ and the smallest value point $b$ of $B$ are equal-value, namely, $\underline{a} = \underline{b}$. Let $\underline{a} = \underline{b} = r$ ($r \in \mathbb{R}$), $a = r_A$, $b = r_B$, then, $r_A \neq r_B$. This implies that there are two equal-value unequal-series points instead of one point in each position in $W$.

**Theorem 3.2** *For $W$ defined by the disordered cut, the Cantor-type continuity and the Poincaré-type continuous are equivalent to each other.*

**Proof** Necessity. Suppose otherwise, i.e., that there is at least a discontinuity point to make $W$ that is not ordered-type continuous. Let $c$ be the discontinuity point for $W$, i.e., $c \notin W$. We cut $W$ into two nonempty parts $A$ and $B$ in the position $\underline{c}$, $A \cup B = W$, $A \cap B = \varnothing$. Since $W$ is Poincaré-type continuous, there are the largest value point $a$ of $A$, and the smallest value point $b$ of $B$. Since $\underline{W} = \underline{A} \cup \underline{B}$ is ordered, then $\underline{a} < \underline{c} < \underline{b}$, and $\underline{A} \cap \underline{B} = \varnothing$, namely, there are $u \in A$ and $v \in B$, then $\underline{u} = \underline{v}$. By Definition 2.2, $A$ and $B$ are not in contact, which contradicts the Poincaré-type continuity of $W$.

Sufficiency. Let $A$ and $B$ be any two parts of $W$, $A \cup B = W$, $A \neq \varnothing$, $B \neq \varnothing$, and $A \cap B = \varnothing$. By the disordered cut, here are the largest value point $a$ of $A$, and the smallest value point $b$ of $B$. Since $W$ is Cantor-type continuous, the value source $\underline{W} = \underline{A} \cup \underline{B}$ is complete. Then we have $\rho(\underline{A}, \underline{B}) = 0$, or $\rho(\underline{a}, \underline{b}) = 0$, i.e., $\underline{a} = \underline{b}$. This implies $A$ and $B$ are in contact; and thus $W$ is Poincaré-type continuous. □

Theorem 3.2 shows that $W$ can simultaneously meet ideas of intuitionists (such as Poincaré) and other mathematicians (such as Cantor) about the continuity of continuum. This implies that the differences in the continuity of continuum are well eliminated in $W$.

**4 Application**

To further discuss application of the set $W$ constructed by using the original idea of Dedekind's cut, we take the vertical motion of an "upcast" small ball as an example. Obviously, the motion trajectory of the ball (denoted by $L$) may be divided into the upcast motion phase and the downward motion phase (denoted by $L_{up}$ and $L_{down}$ respectively). Denoting the highest position of the motion by $p$, the question

(denoted by Q*) naturally arises: does the highest position $p$ belong to the upcast motion phase $L_{up}$ or the downward motion phase $L_{down}$?

For the question Q*, O1 forces the position $p$ can only belong to one of two motion phase $L_{up}$ and $L_{down}$, but it is uncertainty that $p$ belongs to actually which motion phase. This is because, the mathematical model describing motion trajectory of an object that satisfies Order axioms, provides only the spatial information (namely, the positions) of the object without the moving information from one position to another. Thus the question Q* is difficult to answer in the categories of traditional mathematic field satisfying O1.

As two different phases of the motion trajectory $L$, $L_{up}$ and $L_{down}$ satisfy $\rho(L_{up} \cap L_{down}) = 0$ in the position $p$, and are nonempty bounded closed sets because of their end points and start points. This is obviously forbidden by O1. So we use the disordered set $W$ as a mathematical model to describe the motion trajectory of the small ball. For the motion trajectory, there are two points $p_{up}$, $p_{down}$ in the position $p$, where $p_{up}$, $p_{down}$ are the end point of the upcast motion phase and the start point of the downward motion phase, respectively. Note that $p_{up}$, $p_{down}$ are equal-value unequal-series, i.e., $p_{up} \neq p_{down}$ but $\underline{p}_{up} = \underline{p}_{down} = p$. By Definition 3.1, $L_{up}$ and $L_{down}$ are in contact in the position $p$. In fact, two parts $L_1$, $L_2$ obtained by cutting the motion trajectory $L$ in any position are apparently in contact, which implies $L$ is in contact everywhere.

Now, we can answer the question Q* as follows: The highest position $p$ belongs to the upcast motion phase and the downward motion phase. This is because that there are two equal-value unequal-series points in the highest position $p$, i.e., the end point of the upcast motion phase $p_{up}$ and the start point of the downward motion phase $p_{down}$.

The answer to the question Q* reveals implicitly the continuously changing process of an object during moving. The motion trajectory of an object is traditionally described as a finite complete linear ordered set, such as the real numbers $\mathbb{R}$. This inevitably causes the jumping motion of an object from one position to another, because all elements in an ordered (or even complete) set are external to each other. The answer suggests that the continuously changing process of an object should not be ordered, but disordered, which is why we use $W$ as a mathematical model to describe a motion trajectory. For the vertical motion of an "upcast" small ball, from the aspect of "spatial position (value)", the small ball is stationary in the highest position $p$ because $\underline{p}_{up} = \underline{p}_{down} = p$, namely the spatial position produces no change, but from the aspect of "series", the small ball is moving because the series are changing (i.e., "up" → "down") in the highest position $p$. Actually, the same conclusion is obtained for any position $x \in \underline{L}$ of the motion trajectory $L$. Correspondingly, for any series $y \in L_{\#}$ of the motion trajectory $L$, we divide $L$ into the "before phase" $S$ and the "after phase" $T$ (where $S \cup T = L$, $S \cap T = \varnothing$). Then in the series $y$ there exist two unequal-value equal-series points $s \in S$, $t \in T$, i.e., $\underline{s} \neq \underline{t}$, $s_{\#} = t_{\#} = y$. It follows that, from the aspect of "spatial position", the small ball is moving on the series $y$ because the positions are changing (i.e., $\underline{s} \to \underline{t}$), and from the aspect of

"series", the small ball is stationary because the series produces no change (i.e., $s_\# = t_\# = y$) on the series $y$. In short, the motion of an object is realized by appearing alternatively of value changes and series changes of elements (points), which shows the continuous moving of an object on the motion trajectory.

**Acknowledgements**   The author would like to thank Prof. Caisheng Chen & Hongguang Sun for discussion and inspiring comments.